\documentclass[preprint2]{aastex}

\usepackage{graphicx}
\usepackage{color}
\usepackage{soul} 

\shorttitle{The Massive CO White Dwarf in RS Ophiuchi}
\shortauthors{Miko{\l}ajewska and Shara}

\begin{document}


\title{The Massive CO White Dwarf in the Symbiotic Recurrent Nova RS Ophiuchi}
\author{Joanna Miko{\l}ajewska}
\affil{N. Copernicus Astronomical Center, Polish Academy of Sciences, Bartycka 18, 00--716 Warsaw, Poland}
\email{mikolaj@camk.edu.pl}
\and 
\author{Michael M. Shara}
\affil{Department of Astrophysics, American Museum of Natural History, Central Park West at 79th Street, New York, NY 10024, USA}
\email{mshara@amnh.org}



\begin{abstract}

If accreting white dwarfs (WD) in binary systems are to produce type Ia supernovae (SNIa), they must
grow to nearly the Chandrasekhar mass and ignite carbon burning.
Proving conclusively that a WD has grown substantially since its birth is a challenging task.
Slow accretion of hydrogen inevitably leads to the erosion, rather than the growth of WDs. Rapid hydrogen accretion {\it does} lead to growth of
a helium layer, due to both decreased degeneracy and the inhibition of mixing of the accreted hydrogen with the underlying WD.
However, until recently, simulations of helium-accreting WDs all claimed to show the explosive ejection of a helium envelope once it exceeded $\sim 10^{-1}\, \rm M_{\sun}$.
Because CO WDs cannot be born with masses in excess of $\sim 1.1\, \rm M_{\sun}$, any such object, in excess of $\sim 1.2\, \rm M_{\sun}$, {\it must} have grown substantially. We demonstrate that the WD in the symbiotic nova RS Oph is in the mass range 1.2-1.4\,M$_{\sun}$. We compare UV spectra of RS Oph with those of novae with ONe WDs, and with novae erupting on CO WDs. The RS Oph WD is clearly made of CO, demonstrating that it has grown substantially since birth. It is a prime candidate to eventually produce an SNIa.
 
\end{abstract}

\keywords{white dwarfs --- binaries: symbiotic ---  novae, cataclysmic variables --- stars: supernovae: general --- stars: individual (RS Oph)}

\section{Introduction}

Classical novae are all cataclysmic binary stars, wherein a WD accretes a hydrogen-rich envelope from its Roche-lobe filling companion, or from the wind of a nearby giant. Theory \citep{shara1981, fujimoto1982} 
and detailed simulations \citep{yaron2005, townsley2004} predict that once the pressure at the base of the accreted envelope (which has a mass of $\sim 10^{-5}$-$10^{-6} \, \rm M_{\sun}$) of the WD exceeds a critical value, a thermonuclear runaway (TNR) will occur in the degenerate layer of accreted hydrogen. The TNR causes the rapid rise of the WD's luminosity to $\sim 10^{5} \, \rm L_{\sun}$ or more, and the high-speed ejection of the accreted envelope in a classical nova explosion \citep{starrfield1974, prialnik1979}. Most novae must recur on timescales of $\sim 10^{5} $ years \citep{ford1978}, but there are $\sim 30$ objects, in the Milky Way \citep{schaefer2010}, M31 \citep{shafter2015} and the LMC \citep{bode2016} -- recurrent novae (RNe) - that are observed to erupt at least twice per century. RNe must contain WDs close to the Chandrasekhar limit, and they must be accreting at very high rates $\sim 1$-$7 \times 10^{-7} \, \rm M{_\sun}\,yr^{-1}$ \citep{kato1991, yaron2005, hillman2016} in order to build critical-mass envelopes quickly enough to erupt so frequently.

Because the WDs in RNe must be massive, and because they must be accreting at very high rates, it is natural to ask if these WDs can grow secularly in mass to approach the Chandrasekhar limit \citep{whelan1973}, ignite carbon burning and produce SNIa \citep{hillebrandt2000}. This "single-degenerate" (SD) scenario \citep{ruiz1995, branch1995} has long been considered to lead to sub-Chandrasekhar mass WD supernovae, produced by helium flashes on WDs which manage to burn and retain their rapidly accreted hydrogen \citep{rappaport1994}. Because accreted helium envelopes on WDs cannot exceed $\sim 10^{-1} \, \rm M_{\sun}$, it has seemed that the secular growth of moderate mass carbon-oxygen (CO) WDs to the Chandrasekhar limit is blocked. This is puzzling because, as we will describe below, there exist a few CO WDs in close binaries that must be remarkably close to the Chandrasekhar limit.

\citet{hillman2016} have recently modeled the rapid accretion of hydrogen onto CO WDs, leading to the buildup $\sim 0.1 \, \rm M_{\sun}$ shells of helium. Previous similar studies simulated just one or a few of the resulting, successive helium flashes \citep{idan2013, newsham2014, piersanti2014}, concluding that much of the helium must be ejected in those flashes. \citet{hillman2016} showed that dozens of helium flashes heated the WD enough to relieve the degeneracy  in its outer layers. This led to much less violent helium flashes which produce carbon ash, and the retention of much of that ash. This mechanism is capable of growing CO WDs, in principle, to the Chandrasekhar mass. 

A prime example of a frequently erupting and therefore massive WD \citep{wolf2013, kato2014} is the symbiotic binary star RS Oph, which has erupted as a RN in 1898, 1933, 1958, 1967, 1985 and 2006. Eruptions may also have occurred in 1907 and 1945 when RS Oph was aligned with the Sun. 
The supersoft phase exhibited by RS Oph also suggests that its WD is massive \citep{osborne2011, wolf2013}.
RS Oph contains an M0-2 III mass donor \citep{dobrzycka1996, anupama1999} in a binary with a 453.6 day orbital period \citep{brandi2009}. 

\citet{brandi2009} have presented compelling evidence for a massive WD in RS Oph. 
Seventy spectra, obtained over the decade 1998-2008 determined the orbital period to be 453.6 days, and
a mass ratio $q = M_{\rm g}/M_{\rm h} = 0.59 \pm 0.05$.  The most likely orbital solution for the WD mass is $M_{\rm h} = 1.2$-$1.4\, \rm M_{\sun}$, the red giant mass is $M_{\rm g} = 0.68$-$0.80\, \rm M_{\sun}$,  while the orbit inclination is $i = 49\degr$-52$\degr$. We are unaware of any published evidence or claims linking the WD in RS Oph with either a CO or ONe WD. 

Since CO WDs are not born with masses in excess of 1.1\,M$_{\sun}$ \citep{ritossa1996}, demonstrating that the WD in RS Oph is of CO composition would be strong evidence that it has grown considerably in mass since its birth. 
If such growth continues towards the Chandrasekhar mass then RS Oph will end its life as a SNIa. In the following section we describe the compelling evidence that RS Oph is, indeed, a CO WD.

The goal of this paper is to  show that the WD in RS Oph is {\it not} an oxygen-neon (ONe) WD.  While rare, ONe WDs can be born near the Chandrasekhar mass limit. CO WDs do not exceed $\sim 1.1 \, \rm M_{\sun}$ at birth. Demonstrating that the RS Oph WD is, in fact, close to the Chandrasekhar mass {\it and} that it is a CO WD, would be equivalent to demonstrating that it has grown significantly in mass since its birth. If this trend continues then RS Oph will eventually produce an SNIa. As important, it would signal that the progenitors of at least some fraction of SNIa are RS Oph-like systems, i.e. symbiotic stars.

The nature of the underlying WD is usually reflected in the abundances of the novae ejecta, and so, in their linear spectra. 
To determine the abundances of nova ejecta, a majority of modern studies use the CLOUDY photoionization code \citep[][and references therein]{ferland1998} to fit emission line spectra, preferably covering the UV-optical-near infrared range, and collected during the nebular phase when the gas become optically thin \citep[e.g.][and references therein]{schwarz2007}. 
The early nebular phase is the most suitable to study the nova chemistry because the emission lines are strong, the typical gas temperature is around 10000-15000 K and reasonably well known atomic  parameters are available, and possible contamination of the ejecta by swept up interstellar material can be ignored \citep[e.g.][and references therein]{snijders1989}. 

Extensive observational and theoretical studies of the abundances in nova ejecta  allowed to distinguish two different types of classical novae. A majority of them show large CNO overabundances and nearly solar abundances of heavier elements. There is, however, a smaller group with large overabundances (by an order of magnitude and more) of Ne, Mg, Al  and other heavier elements \citep[e.g.][]{livio1994}. These two types, called CO and ONe novae, respectively, develop very different spectra once they enter the nebular phase. The latter show strong  emission lines of  Ne, Mg , Al, and Si, in addition to strong H, He and CNO lines usually observed in novae \citep[e.g.][]{williams1985}. 
The difference between the CO and ONe nova is particularly remarkable in the ultraviolet region where some elements (e.g. Mg, Al, Ne and Si) and many stages of ionization of CNO have strong lines which makes the ultraviolet spectra fundamental to get  a fairly complete set of abundances. 

\begin{figure}
\centerline{\includegraphics[width=0.98\columnwidth]{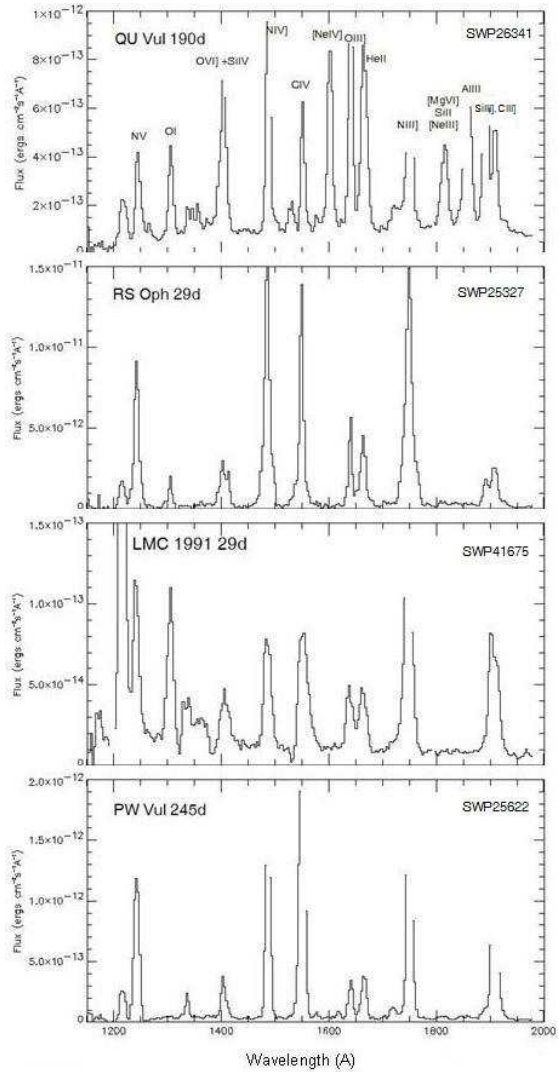}}
\caption{Comparison of IUE low resolution spectrum of RS Oph, the ONe nova QU Vul, and two CO novae LMC 1991 (very fast) and PW Vul. The spectra were taken during similar phases of the nova evolution. Breaks in the spectra are due to saturation in the IUE detector.}

\label{sp1}
\end{figure}
\begin{figure}
\centerline{\includegraphics[width=0.98\columnwidth]{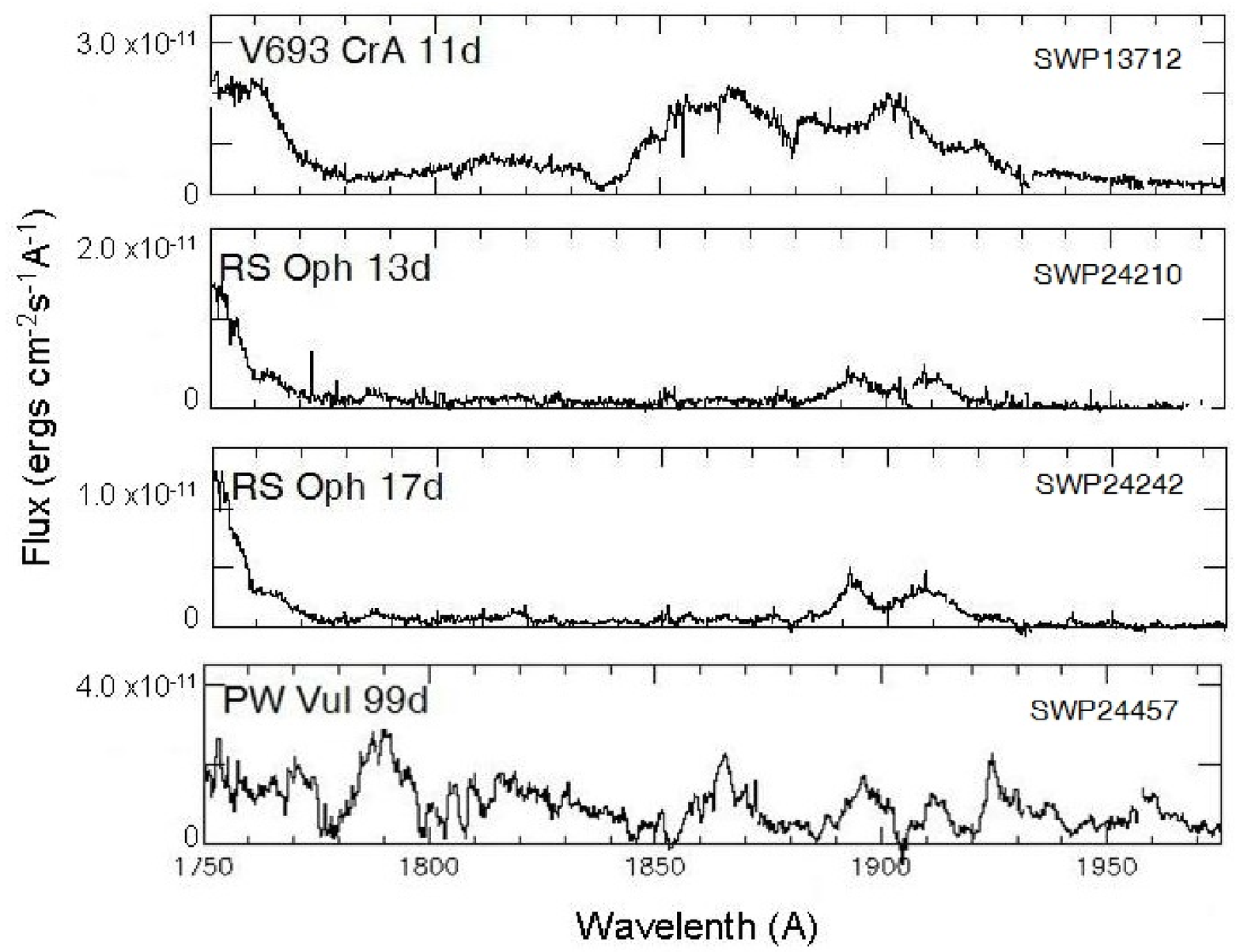}}
\caption{High resolution IUE spectra in the region of the \ion{Al}{3}\,1855,1863 resonance lines and \ion{Si}{3}]\,1882, \ion{C}{3}]\,1907 intercombination lines for RS Oph (day 13 and day 17 of the 1985 outburst), the ONe nova V693 CrA and the CO nova PW Vul. The spectra were taken in roughly similar phase of the nova outburst.}
\label{sp2}
\end{figure}

In Section 2 we describe and analyse the UV spectroscopy of RS Oph, and use it to demonstrate that the WD in this binary is composed of CO, and not oxygen and neon. We summarise our conclusions in Section 3.

\section{UV and optical spectroscopy}\label{sect2} 

The 1985 outburst of RS Oph was extensively observed at both low and high resolution by the IUE satellite. The high resolution data were analysed and discussed in detail by \citet{shore1996}, and the main results highlightened in the review presented by \citet{shore2008} during the Keele meeting on RS Ophiuchi  (2006) and the recurrent nova phenomenon. In particular, the high resolution ultraviolet (as well as optical) emission line profiles showed two separate contributors: a broad-line component emitted from the high-velocity ejecta, and a narrow-line component originating from the red giant wind ionized by the UV flux produced in the nova explosion and by radiation from the shock produced when the ejecta passed though the wind. The situation is thus more complicated than in a typical classical (non symbiotic) nova where the bulk of line emission originates from the nova ejecta. To address properly the question whether RS Oph is a CO or ONe nova one must be confident that the diagnostic is based on features and epochs in the nova evolution dominated by the nova ejecta and not the red giant wind.

\citet{shore1996} demonstrated that the broad-line component behaves as would be expected from an explosive mass ejection. This nova ejecta phase lasted $\sim$ 2 months, i.e. until $\sim$ MJD 46160. In practice, however, the best data for our purpose was the first month after the outburst when the ejecta dominated the emission line profiles, and the contribution from the red giant wind component to the line profiles was negligible (see figs 10 and 11, and discussion in Shore et al. 1996; also Shore 2008).

Fig.~\ref{sp1} presents the low resolution spectrum of RS Oph (middle) taken with the International Ultraviolet Explorer (IUE) on 26 Feb 1985, 29 days after the outburst (when the contribution from the ejecta was dominant) .
We also show the low resolution IUE spectra of the ONe nova  QU Vul (top) and two CO novae: PW Vul, 
and the extremely fast nova LMC 1991, which declined by 3 mag from maximum light in just 8 days ($t_3=8$ days) . 
All spectra correspond to the epoch when the emission line spectrum was strongest, and roughly represent the same phase of nova outburst development. We note that the nova evolution, including spectroscopic characteristics, scales with $t_3$ \citep{cassatella2004, vanlandingham2001}. Here we adopted $t_3=14 $ days \citep{schaefer2010} for RS Oph, and the values from \citet{cassatella2004}, and \citet{vanlandingham2001} for the CO and ONe novae, respectively.

The difference between the ONe nova and RS Oph as well as the CO novae is remarkable.  In particular, QU Vul, in addition to strong CNO and \ion{He}{2} lines, shows very strong  [\ion{Ne}{4}]\,1602 A, the blend at $\sim 1810\, \rm \AA$ made of [\ion{Mg}{6}]\,1806, [\ion{Ne}{3}]\,1815 and \ion{Si}{2}\,1808, and \ion{Al}{3}\,1860. A very strong \ion{Al}{2}]\,2669 line is also present in the long wavelength IUE spectrum (not shown here). Similar UV spectra were observed for other ONe novae, e.g. V693 CrA \citep[][their fig. 1]{williams1985} or V382 Vel \citep[][their fig. 6]{shore2003}. 
All these features are absent in both the CO novae LMC 1991 and PW Vul, as well as  in any other of seven CO novae observed with the IUE \citep{cassatella2005} 
including the most intensively monitored V1668 Cyg \citep{stickland1981} and in RS Oph at any epoch. Thus the similarity between RS Oph and CO novae is evident.
These differences in emission line spectrum appearance reflect the high overabundance of Ne, Mg and Al - over one order of magnitude - with respect to solar, in QU Vul (e.g. Schwarz 2002) and other ONe novae \citep{schwarz2007} as predicted for such novae by theoretical models \citep[e.g.][and references therein]{prialnik1995, jose1999, yaron2005}.  

Fig.~\ref{sp2} shows the high resolution IUE spectra in the region of the \ion{Al}{3} 1855,1863 resonance lines and \ion{Si}{2}I]\,1882, \ion{C}{3}]\,1907 intercombination lines for RS Oph (day 13 and day 17 of the 1985 outburst), another ONe nova V693 CrA and the CO nova PW Vul, shortly after a strong emission-line spectrum appeared. Again, there is a clear difference between RS Oph and PW Vul, and the ONe nova. In particular, the resonance \ion{Al}{3} lines are strong in V693 CrA, and absent in RS Oph and PW Vul. In addition, they show broad P Cyg absorption components. Similar resonance-line profiles, including that of \ion{Al}{3}, were observed in other ONe, e.g. in IUE spectra of V1974 Cyg \citep[][their fig. 3]{cassatella2004}, and  HST/STIS spectra of V382 Vel \citep[][their fig. 7]{shore2003}.
Such broad P Cyg absorption components on the UV resonance lines were not seen in any of the CO novae observed with the IUE \citep[e.g.][]{cassatella2004}, and they are a phenomenological characteristic of the ONe novae \citep{shore1994}. 

The evidence for a CO nova in RS Oph is very clear, especially if one keeps in mind that whereas the contribution from the ionized red giant wind can change the lines ratios (C:N:O), it cannot account  for the absence of any emission lines typical for ONe nova. Morever, differential extinction caused by absorption lines (the so called iron curtain) in the neutral/partly ionized red giant wind (which can be a serious problem in all symbiotic stars including the recurrent novae; e.g. Shore \& Aufdenberg 1993) cannot account for the lack of [\ion{Ne}{4}]\,1602, the 1810 blend, and \ion{Al}{3}\,1860 because all of these features fall in a spectral region relatively free of other lines (see fig. 7 of Shore \& Aufdenberg). One should rather expect that the \ion{C}{4}, \ion{O}{3}] and \ion{He}{2} lines will be much more affected. Thus, the presence of strong and broad \ion{C}{4}, \ion{O}{3}], and \ion{He}{2}, while the [\ion{Ne}{4}]\,1602, the 1810 blend, and  \ion{Al}{3}\,1860 were absent provides very robust evidence that the ONe  WD signatures are not hidden in the red giant wind.
In principle, the lack of [\ion{Ne}{4}]\,1602, and the 1810 blend could also be accounted for by the relatively high, $\ga 10^8\, \rm cm^{-3}$, density in the ejecta  (the critical densities are $\sim 5 \times 10^7$, $\sim 1.4 \times 10^8$, and $\sim 4 \times 10^6 - 1.7\times 10^4\, \rm cm^{-3}$, for [\ion{Ne}{4}]\,1602, [\ion{Ne}{3}]\,1815 and [\ion{Mg}{6}]\,1806, respectively). However, the lack of \ion{Al}{3} cannot be explained by this hypothesis.

Unfortunately, the 2006 outburst of RS Oph was not observed in the UV, so there is no information about the presence or absence of the emission lines typical for ONe nova. However \citet{nelson2011} found from modeling the X-ray and UV emission (XMM-Newton Optical Monitor) that the Ne/O ratio is within a factor of 2 of solar (see their table 3), which strongly argues against ejecta polluted with the Ne of an ONe WD. 
\citet{das2015} applied CLOUDY to low resolution optical and near infrared spectra obtained during the second month after the 2006 maximum, and found N/H  enhanced by a factor of 12 and smaller enhancements of He/H ($1.8\times$), Ne/H ($1.5 \times$), Ar/H ($5\times$) and Fe/H ($3\times$) with respect to solar abundances in addition to solar O/H and Al/H and subsolar Si/H. These results, however, are based on 1-2 lines only except for He and Fe, and it is not clear (no line profiles available) whether their spectra are representative for the ejecta. On the other hand, \citet{brandi2009} observed strong narrow-line components and a broad pedestal in all emission lines on high-resolution optical spectra taken 53-54 days after the 2006 outburst maximum which means that by then the red giant wind contribution to the line emission became significant.

To verify our assumption that the emission line spectrum of RS Oph during the  first month following the nova outburst is indeed dominated by the ejecta, we estimated the relative CNO abundances from fluxes of collisionally excited lines of  CNO. \citet{nussbaumer1988} argued that for high nebular densities, $\ga 10^6\, \rm cm^{-3}$, \ion{C}{3}, \ion{C}{4}, \ion{N}{3}, \ion{N}{4} and \ion{O}{3} are emitted in a common region, and used the UV multiplets \ion{C}{3}]\,1908, \ion{C}{4}\,1549, \ion{N}{3}]\,1749, \ion{N}{4}]\,1486 and \ion{O}{3}]\,1664 to derive N/O and C/N abundances. On the other hand, \citet{williams1981} derived N/O and C/N from  F(\ion{N}{3}]1749)/F(\ion{O}{3}]1664])  and F(\ion{C}{4}1549)/F(\ion{N}{4}]1486) ratios, respectively. Using the broad line fluxes measured on high resolution spectra taken between day 17 and day 30 by \citet[][their table 5]{shore1996} we estimate the average N/O= 1.6-1.9 and C/N=0.1 using Williams et al's method, and N/O=2.8 and C/N=0.07 with Nussbaumer et al.'s approach.  The enhancement of N/O by a factor 10-20 with respect to solar, and depletion of C/N  to 1/40 solar, implies enrichment of N and depletion of C in CNO cycle burning as expected for the nova ejecta. 
In particular, this is evidence of a TNR in material that was enriched in CNO isotopes by an order of magnitude relative to solar material. RS Oph erupts far too frequently for diffusion of accreted hydrogen to penetrate into, and mix with the underlying WD \citep{yaron2005}. Rather, rotationally-induced instabilities or convective boundary mixing during the flash must be at work \citep{glasner2012, casanova2016, denissenkov2017}.

Additionally, the very high N/O is consistent with an outburst on a very massive WD \citep{livio1994, vanlandingham1999}. Similar high N/O abundances were found for the fastest  CO (including LMC 1991) and  ONe novae \citep[e.g.][]{livio1994, schwarz2001, schwarz2007}.  Although N is also enhanced in the atmosphere of the red giant component of RS Oph ([N/H]=0.6$\pm$0.3),  its moderate C/N=0.16  \citep{pavlenko2008} indicates this N enhancement is not as strong as that in the nova ejecta. Systematic studies of photospheric abundances of a few dozens symbiotic giants revealed N/O $\la$ 0.5 and C/N $\ga$ 0.2, and they are consistent with moderate enhancement of N and depletion of C as expected for their evolutionary stage \citep{galan2016, galan2017}. 
   
In summary, the comparison of the IUE spectra of RS Oph and those of classical novae definitely rules out RS Oph as a possible ONe nova, and  demonstrates that the nova outburst occurred on a very massive CO WD.

Occasional claims (none of which are published anywhere) that RS Oph could be a Ne-rich nova have all been  based on the fact that some authors reported optical [\ion{Ne}{3}]\,3869/[\ion{O}{3}]\,5007 greater than unity. In particular, the observed relative flux ratios measured in April 1985 in \citet{wallerstein1986} are consistent with [\ion{Ne}{3}]\,3869/[\ion{O}{3}]\,5007=1.1.  Relative line intensities measured in April 2006 by Iijima (2009) also indicated [\ion{Ne}{3}]\,3869/[\ion{O}{3}]\,5007 up to $\sim$2. Some of the SMARTS spectra\footnote{http://www.astro.sunysb.edu/fwalter/SMARTS/NovaAtlas/rsoph/spec/} also show [\ion{Ne}{3}]\,3869 stronger or comparable to [\ion{O}{3}]\,5007. 
On the other hand, the optical emission line fluxes derived from two absolutely calibrated spectra taken 33 and 126 days after the 1985 maximum give  [\ion{Ne}{3}]\,3869/[\ion{O}{3}]\,5007 of 0.5, and 0.34, respectively  \citep{dobrzycka1996}. These values are typical for symbiotic stars, including symbiotic novae. Moreover, classical symbiotic stars occasionally show  [\ion{Ne}{3}]\,3869/[\ion{O}{3}]\,5007 $\sim$1, e.g., CI Cyg \citep{mik1985} and Z And \citep{mik1996}, however, their radial velocity curves demonstrate that they {\it cannot} contain massive ONe WDs. In addition, in RS Oph,  the emission lines contain variable contributions from both the ejecta and red giant wind, and the wind component dominates the profile as the nova brightness declines \citep{shore1996}. So, without detailed analysis of the line profiles, one cannot draw conclusions about the origin (nova ejecta or red giant wind) of the [\ion{Ne}{3}] and [\ion{O}{3}] lines. In particular, the  [\ion{Ne}{3}]\,3869/[\ion{O}{3}]\,5007 ratio based on line peak intensities (like those given by Iijima) can be very misleading because these probe the narrow line (wind) component, rather than the broad line (ejecta) component which alone are diagnostic of the underlying WD.

\begin{figure}
\centerline{\includegraphics[width=0.98\columnwidth]{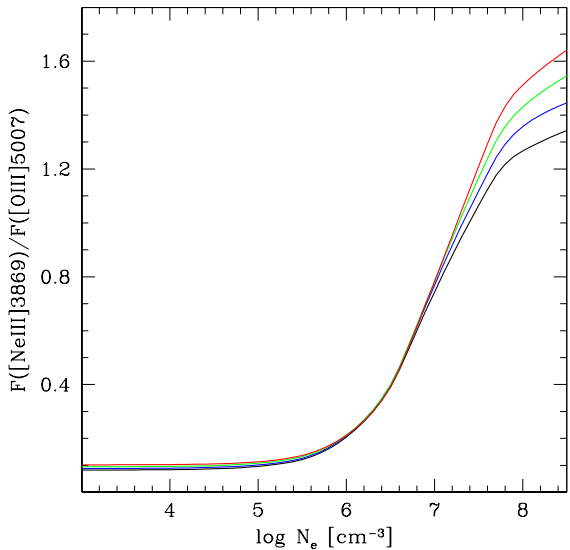}}
\caption{The  [\ion{Ne}{3}]\,3869/[\ion{O}{3}]\,5007 flux ratio vs. electron density. The lines correspond to different Te: 10000 K (the lowest line), 12000 K, 15000 K, and 20000 K (the uppermost).}
\label{ne2o}
\end{figure}

These high [\ion{Ne}{3}]\,3869/[\ion{O}{3}]\,5007 ratios in symbiotic stars result from the relatively high electron densities, $\sim 10^7\, \rm cm^{-3}$ and higher,  in the line formation region(s) in the ionized portion of the red giant wind. The densities there are  generally comparable to or greater than  the critical density ($\sim 10^6\, \rm cm^{-3}$ for [\ion{O}{3}]\,5007, and  $\sim 10^7\, \rm cm^{-3}$ for [\ion{Ne}{3}]\,3869). Fig.~\ref{ne2o} shows the effect of density on the observed [\ion{Ne}{3}]\,3869/[\ion{O}{3}]\,5007 flux ratio.  To calculate the [\ion{Ne}{3}]/[\ion{O}{3}] flux ratio solar Ne/O was assumed and the formula from \citet{ferland1978} was used. 
The plot demonstrates that there is no problem to have this line ratio above 1 with solar Ne/O abundance if the electron density is above the critical value(s), and so, the line ratio cannot be used to derive abundance at high densities (say $\ga 10^6\, \rm cm^{-3}$), without including  collisional de-excitation effects. 

Recently, \citet{mason2011} derived a very high Ne/O for U Sco from the [\ion{Ne}{3}]\,3869/[\ion{O}{3}]\,5007 line ratio observed during its 2011 nova ouburst. However, she neglected the effects of collisions on the line formation. Fig.~\ref{ne2o} shows that for the electron density $\ga 5 \times 10^6 - 10^7\, \rm cm^{-3}$ derived for U Sco \citep{mason2011}, the observed [\ion{Ne}{3}]\,3869/[\ion{O}{3}]\,5007=0.2-0.4 is consistent with solar Ne/O abundance. Mason (2013) corrected this error. There is also no evidence for any [\ion{Ne}{4}], \ion{Al}{3}, and other  lines characteristic of an ONe nova in the IUE spectra collected during the previous outburst of U Sco. So, in the case of U Sco, there is also no indication for an ONe WD.

Both in summary, and in answer to the referee's excellent question "Why did no one notice this before? The 1985 IUE spectra are over 30 years old", we respond:
It is the {\it combination} of {\it all} of the following five points: 
\begin{enumerate}
\item the realization over the past two decades that a massive CO WD is a demonstration that a WD mass {\it can} grow substantially;
\item  a very recent (2016) theoretical framework that explains how such growth can occur without ejecting all accreted mass;
\item  a 2006 measurement of the (small) ejected mass in the 2006 RS Oph outburst;
\item the rather recent (circa 2009) determination that the WD in RS Oph is, indeed, massive; and 
\item the rigorous use of spectroscopic tools (post 2013) to demonstrate that the RS Oph IUE spectrum rules out RS Oph as an ONe nova,
\end{enumerate}
that together finally allow the results of this paper to emerge.

\section{Conclusions}\label{conclusions}

There is strong evidence that the white dwarf in RS Oph is massive. The spectroscopic orbits are derived for both components, and the WD mass is 1.2-1.4\,M$_{\sun}$ \citep{brandi2009}. A very massive WD, 1.3\,M$_{\sun}$, is also required by the outburst characteristics \citep[e.g.][]{hachisu2001, yaron2005, hillman2016}.  

The UV spectroscopic characteristics, and in particular the lack of {\it any} indication of an ONe WD, indicate that RS Oph hosts a massive CO WD. Since the maximum mass of a CO WD resulting from stellar evolution does not exceed $\sim 1-1.1\, \rm M_{\sun}$ \citep[e.g.][and references therein]{marigo2013}, we can conclude that the WD  in RS Oph must had grown to its large present value due to accretion.
The mass of the nova shell ejected in the 2006 outburst of RS Oph was only $\sim 10^{-7}\, \rm M_{\sun}$ \citep{sokoloski2006}  which means that the WD retains most of its accreted envelope, and so its mass continues to grow.
Our ultimate conclusion is that RS Oph contains a very massive CO WD, growing in mass, which eventually should explode as a SN Ia.


\section{Acknowledgments}

This study has been supported in part by the Polish NCN grant
DEC-2013/10/M/ST9/00086.
We acknowledge very helpful discussions with F. Patat, Ph. Podsiadlowski, and R. Williams. 
This research has made use of the Mikulski Archive for Space Telescopes (MAST), and data from the SMARTS telescope at CTIO.




\end{document}